\documentclass[aps,amsmath,amssymb,reprint,superscriptaddress,preprintnumbers]{revtex4-1}

\usepackage{graphicx}
\usepackage{dcolumn}
\usepackage{bm}

\usepackage[utf8]{inputenc}
\usepackage[T1]{fontenc}
\usepackage{eurosym}
\usepackage{mathptmx}
\usepackage{etoolbox}
\usepackage{amssymb}
\usepackage{array}
\usepackage{booktabs}

\usepackage{xcolor}
\usepackage[normalem]{ulem}
\usepackage{lineno}
\usepackage{physics}
\usepackage{multirow}

\makeatletter
\def\@email#1#2{%
 \endgroup
 \patchcmd{\titleblock@produce}
  {\frontmatter@RRAPformat}
  {\frontmatter@RRAPformat{\produce@RRAP{*#1\href{mailto:#2}{#2}}}\frontmatter@RRAPformat}
  {}{}
}%
\makeatother
\draft 

\begin{document}

\preprint{{\sc Version} \#1}

\title{Hybrid quantum-classical convolutional neural networks to improve molecular protein binding affinity predictions} 

\author{L. Domingo}
\affiliation{Instituto de Ciencias Matemáticas (ICMAT); Campus de Cantoblanco UAM;
Nicolás Cabrera, 13-15; 28049 Madrid (Spain)}
\affiliation{Departamento de Química; Universidad Autónoma de Madrid;
CANTOBLANCO - 28049 Madrid (Spain)}
\affiliation{Grupo de Sistemas Complejos; Universidad Polit\'ecnica de Madrid; 
 28035 Madrid (Spain)}
\affiliation{Ingenii Inc.}

\author{M. Djukic}
\affiliation{Ingenii Inc.}

\author{C. Johnson}
\affiliation{Ingenii Inc.}

\author{F. Borondo}
\affiliation{Instituto de Ciencias Matemáticas (ICMAT); Campus de Cantoblanco UAM;
Nicolás Cabrera, 13-15; 28049 Madrid (Spain)}
\affiliation{Departamento de Química; Universidad Autónoma de Madrid;
CANTOBLANCO - 28049 Madrid (Spain)}

\date{\today}

\begin{abstract}
One of the main challenges in drug discovery is to find molecules that bind specifically and strongly to their target protein while having minimal binding to other proteins. By predicting binding affinity, it is possible to identify the most promising candidates from a large pool of potential compounds, reducing the number of compounds that need to be tested experimentally. Recently, deep learning methods have shown superior performance than traditional computational methods for making accurate predictions on large datasets. However, the complexity and time-consuming nature of these methods have limited their usage and development. Quantum machine learning is an emerging technology that has the potential to improve many classical machine learning algorithms. In this work we present a hybrid quantum-classical convolutional neural network, which is able to reduce by 20\% the complexity of the classical network while maintaining optimal performance in the predictions. Additionally, it results in a significant time savings of up to 40\% in the training process, which means a meaningful speed up of the drug discovery process.
\end{abstract}


\maketitle 


\section{Introduction}
\label{Introduction}

The ability to predict the binding affinity between a potential drug and its target protein is crucial for the success of drug discovery at the early stages. Experimentally determining the binding affinity for a large number of small molecules and their targets is time-consuming and expensive. As a result, computational methods that can predict binding affinity for multiple molecules with high accuracy have become widely used. Machine learning techniques, specifically deep learning methods, have recently gained attention for their ability to improve upon traditional physics-based methods. Unlike traditional machine learning, deep learning can learn directly from the atomic structure of the protein-ligand pair without relying on pre-determined, fixed-length features.\\
\\
A commonly-used deep learning approach for binding affinity prediction is the three-dimensional convolutional neural network (3D CNN) \cite{ATOM, AtomNet, Wave, DeepSite, ScoringCNN, AbsoluteCNN}. These networks represent atoms and their properties in a 3D space and take into account the local 3D structure of molecules and the relationship between atoms. The 3D representations used as input of the 3D CNN are high-dimensional matrices since millions of parameters are required to describe only one data sample. Because of the high-dimensionality of the data, a complex deep learning model is required to uncover all the hidden patterns that can help predict the target value. Training such a model means finding the optimal value for parameters that minimize a loss function. More complex models have more training parameters, requiring longer execution times which limits the exploration of different architectures or hyperparameters. This training process can be heavily accelerated using powerful GPUs. However, GPUs are costly computational resources and may not always be available.\\
\\
The complexity of a machine learning model also affects its generalisation capacity. According to Hoeffding's theorem \cite{MLBook}, models with high complexity require a large amount 
of data to reduce the variance of the model predictions, as stated by Hoeffding's inequality 

\begin{equation}
    E_{out}\leq E_{in} + \mathcal{O}\Big(\sqrt{\frac{K}{N_\text{samples}}}\Big),
    \label{Hoeffding}
\end{equation}
 where $E_{out}$ is the error in the test set, $E_{in}$ is the error in the training set, $K$ is a notion of complexity and $N_\text{samples}$ is the number of data samples. The number of samples should be at least comparable to the complexity of the model to guarantee low errors in the predictions of new data. In some cases, when the test data is similar enough to the training data, a smaller training set can still allow for good performance. Nonetheless, increasing the complexity of a model always increases its chances of producing overfitting, and thus it is convenient to find simpler machine learning models.\\
 \\
 Quantum machine learning (QML) methods have the potential to solve numerical problems exponentially faster than classical methods. Although fault-tolerant quantum computers are still not available, a new trend of quantum algorithms, called noisy intermediate-scale quantum (NISQ) era is devoted to design quantum algorithms that provide quantum advantage with the quantum computers available today. Because of the exponential scaling of the Hilbert space dimension, quantum computers can process large amounts of data with few qubits. For this reason, combining quantum algorithms with machine learning allows to reduce the complexity of the classical machine learning methods, while maintaining its accuracy. In this project, we propose a hybrid quantum-classical 3D CNN, which replaces the first convolutional layer by a quantum circuit, effectively reducing the number of training parameters of the model. The results of our work show that as long as the quantum circuit is properly designed, the hybrid CNN maintains the performance of the classical CNN. Moreover, the hybrid CNN has 20\% less training parameters, and the training times are reduced by 20-40\%, depending on the hardware where they are trained. All the quantum circuits used in this work have been executed using quantum simulation due to the current limitations of quantum hardware. However, we provide performance benchmarks considering different noise models and error probabilities. Our results show that with error probabilities lower than $p=0.01$ and circuits with 300 gates, a common error mitigation algorithm, namely \emph{data regression error mitigation} \cite{DRER}, can accurately mitigate the errors produced by the quantum hardware. For this reason, we believe that hybrid quantum-classical machine learning methods have the potential to speed-up the training process of classical machine learning methods and reduce the computational resources needed to train them.\\
 \\
 The organization of this paper is as follows. In Sect. \ref{Methods} we present the protein-ligand data used for this study, the processing techniques used to prepare the data for the CNN models, and the details of the classic and hybrid CNN approaches. The results and comparison of the classical and hybrid CNNs are presented in Sect.~\ref{Results}. Finally, Sect.~\ref{Discussion} ends the paper by summarizing the main conclusions of the present work, and presenting an outlook for future work.

\section{Methods}
\label{Methods}
This section provides an overview of the classical and quantum machine learning algorithms that were examined in this study. It starts by discussing the PDBBind dataset and the methods used to preprocess the data for use in neural network models. The architecture of a classical 3D CNN is then described. All the processing algorithms and the architecture of the classical 3D CNN are kept equal to the ones in Ref. \cite{ATOM} to support a reproducible and comparable pipeline. Finally, the design of the hybrid quantum-classical CNN is outlined.

\subsection{Data}
  \label{Data}
 The data used for this study is sourced from the PDBBind database \cite{PDBBind}. It contains a collection of protein-ligand biomolecular complexes, manually collected from their associated publications. For each protein-ligand complex, the data files contain information about their 3D morphology, types of bonds between their constituent atoms, together with the binding affinity between the protein and the ligand. All the binding affinities are experimentally obtained by measuring the equilibrium dissociation constant between protein-ligand ($k_d$) and the inhibition constant ($k_I$). Then, the binding affinity is defined as $-\log(\frac{k_d}{k_I})$. Because of its completeness and extension, the PDBBind dataset has recently become a common benchmark for binding affinity prediction with both physics-based and machine learning methods \cite{3DCNNBA, DeepLearningBA, Fingerprints}. The PDBBind dataset is already split into two non-overlapping sets, the general set and the refined set. The refined set is compiled to contain higher-quality complexes based on several filters regarding binding data (e.g complexes with only $IC_{50}$ measurement), crystal structures (e.g low crystal resolution or missing fragments in the complex), as well as the nature of the complexes (e.g ligand-protein covalent bond binding). A subset from the refined set, called \emph{core set} is separated to provide a small, high-quality set for testing purposes. \\
 \\
 The 2020 version of the PDBBind dataset is used for this study. The general set (excluding the refined set) contains 14127 complexes, while the refined set contains 5316 complexes. The core set is significantly smaller, with only 290 data samples.

\subsection{Data processing}
In order to train classical and hybrid CNNs, the raw PDBBind data has to be transformed into an appropriate input format for the convolutional layers. Before reshaping the data into the appropriate format, a common processing protocol was applied, following the same process as Ref. \cite{ATOM,Pafnucy}. Hydrogens were added to all protein-ligand complexes according to each atom's valence. The partial charge of all the bonds is solved based on Amber/GAFF atom types, using Chimera22 with the default settings. This protocol converts the pdb files to Mol2 files. A 3D spatial representation was then used to represent the features of the data. This method uses 3D volume grids to capture the atomic relationships in a voxelized space. That is, each data sample has size $(C,N,N,N)$, where $N$ is the size of each dimension in space, and $C$ is the number of features extracted from the protein-ligand pair. For this project, we set $N=48$, so that each side of the volume space has a size of 48\r{A}, and a voxel size of 1\r{A}. This size allows covering all the pocket region without having too large input sizes for the CNN models. Having set the dimension of the space, 19 features were extracted from each protein-ligand pair ($C=19$). The selected features are the following:
\begin{itemize}
    \item \textbf{Atom type:} One-hot encoding of the elements B, C, N, O, P, S, Se, halogen, or metal.
    \item \textbf{Atom hybridization:} Gives information about the number of $\sigma$ and $\pi$ bonds connecting a particular atom to a neighboring atom. Takes values 1, 2 and 3 for sp1, sp2 and sp3 hybridizations, respectively.
    \item \textbf{Number of heavy atom bonds:} Heavy atoms are all atoms except for hydrogen.
    \item \textbf{Number of bonds with other heteroatoms:} Heteroatoms are those atoms different from hydrogen or carbon.
    \item \textbf{Structural properties:} one-hot encoding of hydrophobic, aromatic, acceptor, donor, and ring properties.
    \item \textbf{Partial charge:} Distribution of charge of an atom  as a result of its chemical environment.
    \item \textbf{Molecule type:} Indicates whether it is a protein atom or a ligand atom (-1 for protein, 1 for ligand).
\end{itemize}

\begin{figure*}
\includegraphics
   [width=0.98\textwidth]{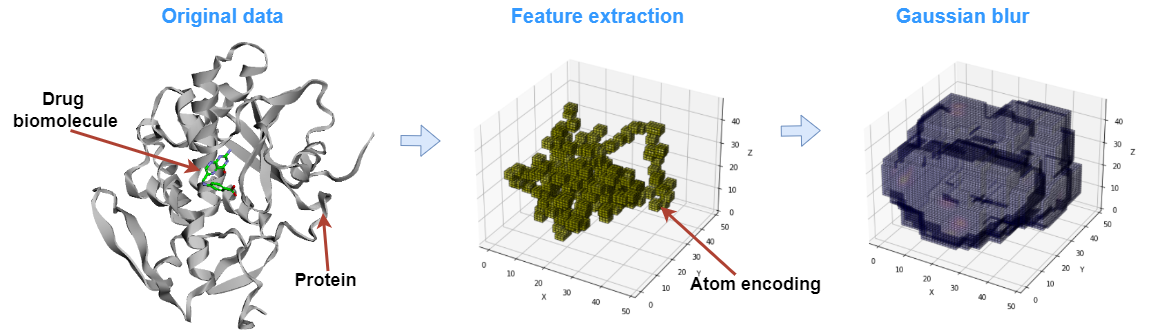}
\caption{Example of a data sample and further processing from the PDBBind dataset. The protein-ligand pair corresponds to the 1br6 sample from the refined set. The first step of the processing is the feature extraction, where 19 features are voxelized in a 3D space. Then, a Gaussian blur is applied to produce a dense representation of the data.}
\label{fig:protein}
\end{figure*}

The feature extraction process was done with the OpenBabel tool (version 3.1.1.1). The Van der Waals radius was used to determine the size of each atom in the voxelized space. In this way, an atom could occupy one or more voxels depending on its Van der Waals radius. For atom collisions, the features were added element-wise. The resulting 3D representations of the features resulted in sparse 3D matrices. Sparse data samples may make the training of neural networks harder since the input samples are too similar to each other, and to a zero-valued sample. Therefore, neural networks can have difficulties differentiating useful information from noise. For this reason, a Gaussian blur with $\sigma=1$ is applied to the voxelized features, populating the neighbouring atoms and thus reducing the number of zero-values voxels. Fig. \ref{fig:protein} shows a representation of the initial protein-ligand pair and the two main processing steps. Note that these 3D volume representations are very high-dimensional since more than 2 million real numbers are needed to represent only one data sample. For this reason, large amounts of data samples and complex neural network models are needed to make accurate predictions without overfitting the training data. \\
\\
The data processing is done independently for each of the datasets considered in this study. Apart from the general, refined and core sets, we further partitioned the general and refined sets into training and validation sets. This split is done to maintain the probability distribution of the binding affinities in both training and validation sets. For this reason, the binding affinities were separated in quintiles. Then for each quintile, we randomly selected 10\% of the data for the validation set, and kept the rest for the training set. In this way, we obtained training and validation sets for both the general and refined sets.

\subsection{Classical CNN}
\label{CNN}
\begin{figure*}
\includegraphics
   [width=0.90\textwidth]{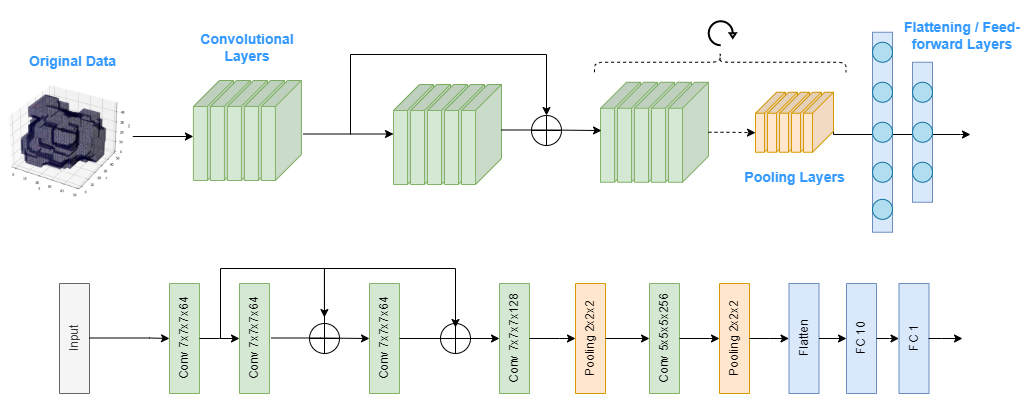}
\caption{(TOP) Schematic representation of the components of a 3D convolutional neural network. (BOTTOM) Architecture of the 3D CNN used for this study, proposed in Ref. \cite{ATOM}.  }
\label{fig:CNN}
\end{figure*}

CNNs have been proven to produce very successful results in deep learning applications. This type of network is specialised in processing high-dimensional data in the form of spatial arrays, such as time series (in 1D), images (in 2D) or volumes (3D). The name stems from the fact that instead of general matrix multiplication it employs a mathematical convolution in at least one of its layers. The output of the convolution is another array that represents some information that was present in the initial array in a very subtle way. In this way, a filter of a convolutional neural network is responsible for detecting one feature of the network input. The kernel matrices are free parameters that must be learned to perform the optimal feature extraction. The convolutional operation is followed by a nonlinear activation function which adds non-linearity to the system. Following the convolutional layers, a pooling layer is added in order to progressively reduce the spatial size of the array. After a series of convolutional and pooling layers, a flattening layer and some feed-forward layers are used to combine the extracted features and predict the final output. \\
\\
A representation of the layers of a 3D CNN is shown in Fig. \ref{fig:CNN} (Top). 3D CNNs have been used for multiple applications such as volume image segmentation \cite{ImageSegm}, medical imaging classification \cite{medicalImages} and human action recognition \cite{HumanAction}. A diagram of the 3D classical CNN used in this work is shown in Fig. \ref{fig:CNN} (Bottom). The architecture is the same as the one proposed in Ref. \cite{ATOM}, again for comparison purposes. The network contains five 3D convolutional layers, with 64,64,64,128 and 256 filters respectively. The kernel size is 7 for all the layers except for the last one, which has kernel 5. The CNN contains two residual connections, as proposed in ResNet \cite{ResNet}, which allow passing gradients to the next layers without a nonlinear activation function. Batch normalization is used after each convolutional layer, and we use Rectified Linear Unit (ReLU) as an activation function. The network contains two pooling layers and two fully-connected layers with 10 and 1 neurons respectively.

\subsection{Hybrid quantum-classical CNN}
\label{QCNN}
\begin{figure*}
\includegraphics[width=0.90\textwidth]{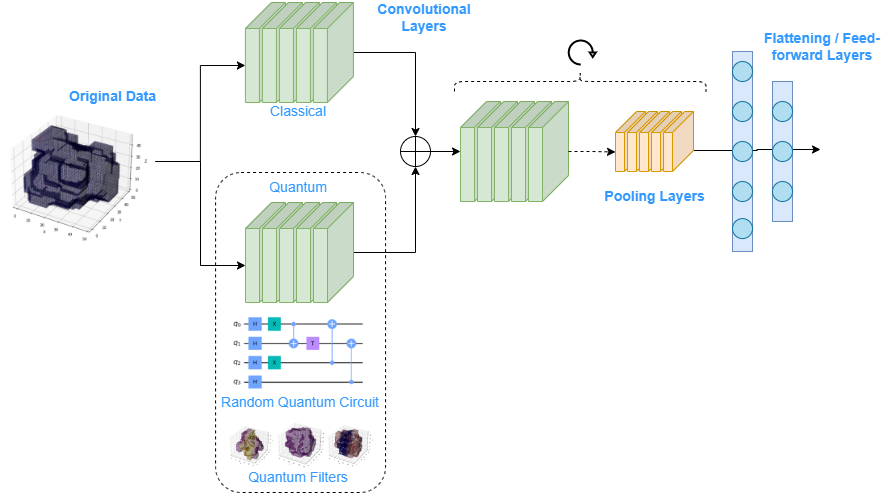}
\caption{Schematic representation of the hybrid quantum-classical (3D) CNN. The original data is processed by both a classical convolutional layer and a quantum convolutional layer. The outputs of both layers are then aggregated. The result is then fed to a set of convolutional and pooling layers, following the same architecture as the classical 3D CNN.}
\label{fig:QCNN}
\end{figure*}

 In this work, we propose a hybrid (quantum-classical 3D) CNN, which is designed to reduce the complexity of the classical 3D CNN, while maintaining its prediction performance. The hybrid CNN replaces the first convolutional layer with a quantum convolutional layer \cite{tutorialQCNN, QCNN, Quanvolutional}. That is, each classical convolutional filter is replaced by a quantum circuit, which acts as a quantum filter. These quantum circuits should have significantly fewer training parameters than the classical convolutional layer, in order to reduce the overall complexity of the network. Each quantum circuit is divided into two blocks: the \emph{data encoding}, which maps the input data into a quantum circuit, and the \emph{quantum transformation}, where quantum operations are applied to retrieve information from the encoded data. The final architecture of the hybrid CNN is depicted in Fig. \ref{fig:QCNN}. The processed protein-ligand data is fed to both a classical and a quantum convolutional layer. The outputs are aggregated by using a residual connection and then fed to the subsequent classical convolutional and pooling layers. The rest of the network is the same as its classical version. With this architecture, the first convolutional layer has been replaced by its quantum counterpart, while leaving the rest of the network unchanged. 
 
\subsubsection{Data encoding}
The quantum convolutional layer aims to extract local features from the input data, just as the classical convolutional layer would do. For this reason, we split the input data into $(n\times n \times n), n \in \mathbb{N}$ blocks and process each block individually. Given a block $B$, the data encoding process converts $B$ to a quantum state $\ket{B}$. Because of the high dimensionality of our data, we need to find a data encoding method that minimizes the number of qubits of the resulting quantum circuit. A suitable encoding should scale logarithmically with the dimension of the blocks. A popular data encoding mechanism that fulfills this property is called amplitude encoding \cite{amplitudeEncoding}, which requires  $\lceil \log_2(n^3) \rceil $ qubits to encode a block. However, the amplitude encoding scheme normalizes each block independently to produce a normalized quantum state. Therefore, the different blocks of the data would have different normalization constants and would not be comparable with each other. For this reason, we decided to choose the Flexible Representation of Quantum Images \cite{FRQI} (FRQI) method, which normalizes the \emph{whole} image before the encoding, avoiding this problem, and uses only $\lceil \log_2(n^3) \rceil + 1$ qubits. FRQI was proposed to provide a normalized quantum state which encodes both the value (colour) of a pixel and its position in an image. Given an image with $\theta = (\theta_0, \theta_1, \cdots, \theta_{2^{n-1}})$ pixels, where the pixels have been normalized such that $\theta_i \in [0, 2\pi), \forall i$, the encoded state is given by Eq. \ref{FRQI}.

\begin{equation}
    \ket{I(\theta)} = \frac{1}{2^n} \sum_{i=0}^{2^{2n}-1} (\cos\theta_i \ket{0} + \sin \theta_i \ket{1}) \otimes \ket{i}
    \label{FRQI}
\end{equation}

where $\ket{i}, i=0,1,\cdots 2^{{2n}-1}$ are the basis computational states. For each $\theta_i$, the FRQI is composed of two parts: $\cos\theta_i \ket{0} + \sin \theta_i \ket{1}$ encodes the color of the pixel, and $\ket{i}$ encodes the position of the pixel in the image. As a simple example, a $(2\times 2)$ image and its representation are displayed in Eq. \ref{FRQI_ex}.

\begin{equation}
\begin{aligned}
    &\begin{array}{|c|c|}
        \hline
        \theta_{0},\text{\small{(00)}} & \theta_{1}, \text{\small{(01)}}  \\
        \hline
        \theta_{2}, \text{\small{(10)}} & \theta_{3}, \text{\small{(11)}} \\
        \hline
    \end{array}, \\
    &\begin{aligned}
        \ket{I}=\frac{1}{2}[ \; & \phantom{+} \left(\cos\theta_{0}\ket{0}+\sin\theta_{0}\ket{1} \right)\otimes\ket{00}&\\
        & + \left(\cos\theta_{1}\ket{0}+\sin\theta_{1}\ket{1} \right)\otimes\ket{01} \\
        & + \left(\cos\theta_{2}\ket{0}+\sin\theta_{2}\ket{1} \right)\otimes\ket{10}\\
        & + \left(\cos\theta_{3}\ket{0}+\sin\theta_{3}\ket{1} \right)\otimes\ket{11} \;]
    \end{aligned}
\end{aligned}
\label{FRQI_ex}
\end{equation}

The state in Eq. \ref{FRQI} is normalized, since

\begin{equation}
    ||\ket{I(\theta)} = \frac{1}{2^n} \sqrt{\sum_{i=0}^{2^{2n}-1} (\cos^2(\theta_i) + \sin^2(\theta_i))} = 1.
\end{equation}
The number of qubits needed to construct the FRQI state increases logarithmically with the number of pixels (angles) of the image, since the dimension of the computational basis increases exponentially with the number of qubits of the Hilbert space. In Ref. \cite{FRQI}, it is proven that the FRQI state can be implemented with simple quantum gates (Hadamard gates, CNOTs and $R_y$ rotations). The number of quantum gates is polynomial with $2^{2n}$, the number of pixels of the image. Even though the FRQI was designed for 2D colour images, the generalization to 3D blocks is straightforward. Let $B$ be a $(n \times n \times n)$ block, with normalized values $(\theta_0, \theta_1, \cdots, \theta_{n^3 - 1}), \theta_i \in [0, 2\pi), \forall i$.   The FRQI state would then be given by Eq. \ref{FRQI3D}.

\begin{equation}
    \ket{B} = \frac{1}{n^3} \sum_{i=0}^{n^3 - 1} (\cos\theta_i \ket{0} + \sin \theta_i \ket{1}) \otimes \ket{i}
    \label{FRQI3D}
\end{equation}

\begin{figure}
\includegraphics[width=0.90\columnwidth]{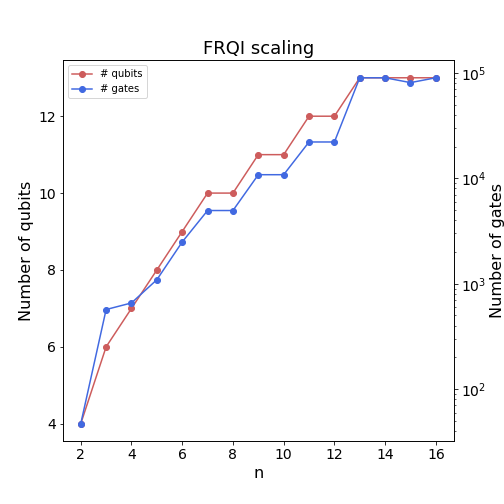}
\caption{Example of scaling of the Flexible Representation of Quantum Images (FRQI). The number of qubits scales logarithmically with the dimension of the block $n^3$. The number of gates of the quantum circuit scales linearly with the dimension of the block $n^3$.  }
\label{fig:FRQI}
\end{figure}

Notice that the only difference between Eq. \ref{FRQI} and Eq. \ref{FRQI3D} is the number of angles of the quantum state. When $n^3$ is a power of 2 (i.e $n^3 = 2^l, l \in \mathbb{N}$), the state in Eq. \ref{FRQI3D} has non-zero components in all the states of the computational basis. Therefore, choosing $n^3$ as a power of 2 mostly exploits the use of the Hilbert space. For this reason, we set $n=4$ for our experiments. Fig. \ref{fig:FRQI} shows an example of the scaling of the number of qubits and the number of gates with the block size $n$. The number of qubits needed for the FRQI encoding is $\lceil \log_2(n^3) \rceil + 1$, so it scales logarithmically with the dimension of the block.\\
\\
On the other hand, we have calculated the number of gates needed to implement the FRQI on a real quantum device. The number of gates depends on the values of the block $\theta_i$. If there are some angles with the same value, the quantum circuit can be compressed to reduce the number of gates. In Ref. \cite{FRQI}, the authors show how the FRQI quantum circuits can be simplified by minimizing boolean expressions. As an example of how the number of gates can scale with the block size, we have considered blocks from our data with the highest mean absolute sum, to ensure that we chose blocks with highly different angles. Fig. \ref{fig:FRQI} shows that the number of gates of a general FRQI encoding scales linearly with the dimension of the block, $n^3$.

\subsubsection{Quantum transformation}
\begin{figure*}
\includegraphics
   [width=0.80\textwidth]{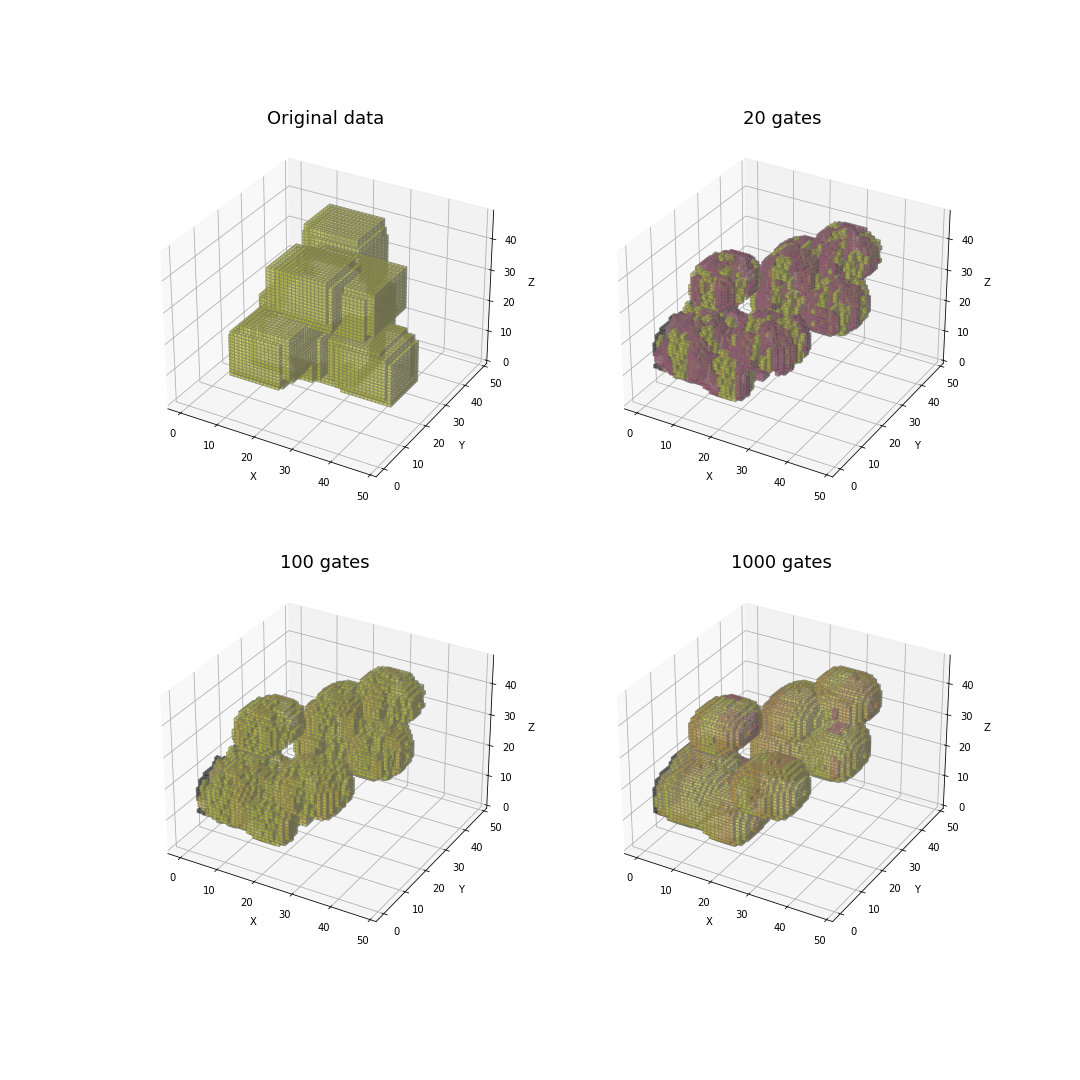}
\caption{Example of an output of the quantum convolutional layer, together with its input. The quantum convolutional layer is composed by a FRQI encoding layer followed by a quantum transformation generated with a random quantum circuit from the G3 family. Examples of the output are given for circuits with 20, 100 and 1000 quantum gates.}
\label{fig:quantum_filters}
\end{figure*}

After the data has been encoded in a quantum circuit, a set of quantum gates is applied to perform the quantum transformation, followed by a set of measurements that convert the data back to a classical representation. In a quantum convolutional layer, the quantum transformation is usually a parameterized quantum circuit (PQC), where the optimal parameters of the circuit need to be learned. However, because of the challenging dimensionality of our data, many quantum complex circuits are needed to process a single data sample. By splitting the image in $(4 \times 4 \times 4)$ blocks, 32832 quantum circuits are required to span the whole sample. The current hardware has limitations not only on the number of qubits and quantum gates but also on the number of quantum circuits that can be executed. For this reason, it is not possible to train the whole neural network model with the current hardware. Another option would be to run the PQC on quantum simulation. Even though the quantum layer would still have fewer training parameters than the classical convolutional layer, the training of quantum neural networks is not prepared to be run on GPUs, as compared with classical convolutional layers. For this reason, even though the hybrid model with PQCs has lower complexity, in our experiments the training time was longer than the classical CNN. \\
\\
The use of quantum reservoirs (QR) is an emerging approach in quantum machine learning (QML), which has provided promising results in multiple tasks \cite{Fujii2021, QRC2,reviewQRC}. It exploits the quantumness of a physical system to extract useful properties of the data that are then used to feed a machine learning model. In gate-based quantum computation, a QR is a \emph{random} quantum circuit 
applied to an initial state, which encodes the input data, followed by measurements of local operators. These measurements are the features extracted by the model, which are then fed to a classical machine learning algorithm to predict the desired output. The main advantage of using QRs is the low complexity of the model, and thus, its easy training strategy. Instead of using PQC and finding its optimal parameters, QRs use carefully selected quantum systems with no training parameters to transform the input data. QRs have been used for temporal tasks (quantum reservoir computing \cite{QRC2,DynamicalIsing}) and also to predict the excited properties of molecular data \cite{quantumchemQRC, Domingo}. The design of the random quantum circuit is crucial to determine the performance of the QML model. Complex quantum circuits are the ones which better exploit the quantum properties of the system, and thus provide useful features for learning the target. In a recent work \cite{Domingo}, it was shown that the majorization principle \cite{majorization_original} is a good indicator of both complexity \cite{majorization} and performance \cite{Domingo} of a QR. That is, the QRs with higher complexity according to the majorization principle are the ones which give better results in the QML tasks. In particular, seven families of quantum circuits, with different complexity, were used as QRs. For a given family, a quantum circuit is built by adding a fixed number of random quantum gates from such family. The G3=\{CNOT,H,T\} family, where CNOT is the controlled-NOT gate, H stands for Hadamard, and T is the $\pi/8$ phase gate, provided the best results when training the QML algorithm. Moreover, the performance of the QR increased with the number of gates of the circuit, until the performance reached its optimal value, and then it remained constant even if the number of gates increased. In this work, the quantum transformation consists of a quantum circuit randomly generated with gates from the G3 family. Then, the qubits are measured in the computational basis, providing the output of the quantum convolutional layer. The hybrid CNN is trained with QRs with 20, 50, 100, 200, 300, 400, 500 and 600 quantum gates. In this way, we can evaluate how the depth of the QR influences the performance of the model. Fig. \ref{fig:quantum_filters} shows an example of the output of the quantum convolutional layer.

We see that with a low number of gates, the quantum layer extracts simpler quantum features than with a higher number of gates. Another widely used QR is the Ising model \cite{Fujii2021, QRC2, quantumchemQRC, OptQRC}. In this case, the quantum circuit performs the time evolution of a quantum state under the random transverse-field Ising Hamiltonian
\begin{equation}
    H_{\text{Ising}} = \sum_{i,j=0}^{N-1} J_{ij} Z_iZ_j + \sum_{i}^{N-1} h_{i} X_i,
\end{equation}
where $X_i$ and $Z_j$ are Pauli operators acting on the site $i, j$-th qubit. The coefficients $J_{ij}$ and $h_i$ are chosen according to Ref. \cite{DynamicalIsing}, which provides a state-of-the-art method to select optimal parameters of the Ising model for quantum reservoir computing. In this case, $J_{ij}$ are sampled from the uniform distribution $U(-J_s/2, J_s/2)$ and $h_i = h$ are constant. The optimal parameters in Ref. \cite{DynamicalIsing} fulfill $h/J_s = 0.1$. The system is evolved until time $T=10$. We will compare the performance of the hybrid CNNs trained with QRs generated from the G3 family as well as the performance of the models with QRs generated from the Ising model. Since the current quantum computers have limitated availability and high access queue times, which limit the number of iterative runs we can do for training, the hybrid CNNs are run using quantum simulation on classical hardware. The code has been optimized using Qiskit and PyTorch, and adapted so that it could be trained on GPUs, just like the classical CNN.

\subsection{Error mitigation}
One of the biggest challenges of the current quantum devices is the presence of noise. They perform noisy quantum operations with limited coherence time, which affects the performance of quantum algorithms. Even though the quantum circuits used for this study are run using quantum simulation, we have also evaluated the performance of the noisy quantum circuits using three different noise models for a small set of samples. The first noise model is the \emph{amplitude damping channel}, which reproduces the effect of energy dissipation, that is, the loss of energy of a quantum state to its environment. The second noise model is described by the \emph{phase damping channel}, which models the loss of quantum information without loss of energy. The last error model is described by the \emph{depolarizing channel}. In this case, a Pauli error $X$, $Y$ or $Z$ occurs with the same probability $p$. For more information about the error models see Ref. \cite{nielsen_chuang}. We perform noisy simulations with error probabilities $p=0.03, 0.01, 0.008, 0.005, 0.003, 0.001$.  Error mitigation methods aim to reduce the noise of the outputs after the quantum algorithm has been executed. In this work, the \emph{data regression error mitigation} (DRER) algorithm is used to mitigate the noise of the quantum circuits. The DRER algorithm trains a machine learning model to correct the errors of noisy quantum circuits. To obtain the training set, random quantum circuits with 300 gates sampled from the G3 family are executed with both noisy and noiseless simulation. Thus, the training set consists of pairs $(X_i,y_i)$ where $X_i$ contains the counts of the noisy distribution and $y_i$ contains the counts of the noiseless distribution. In this case, the machine learning model we used is ridge regression, a regularized linear model which minimizes the mean squared error:

\begin{equation}
    \text{MSE}_R = 
    \frac{1}{N_s} \sum_{i=0}^{N_s} \left[ W \cdot X_i - y_i \right]^2 
    + \alpha ||W||^2
\end{equation}
where $N_s$ is the number of samples in the training set, $W$ is the matrix of the linear model, $\alpha$ is the regularization parameter, and $||\cdot||$ is the $L^2$ norm. The DRER is trained with 1000 samples. Then, the performance is tested with 500 noisy quantum circuits used in the quantum convolutional layer. In this case, the 3D volumetric space is divided in blocks of size $n=8$, leading to quantum circuits of 9 qubits and 300 gates. The DRER algorithm is suitable for this task since, once the machine learning model is trained, it can be used to mitigate multiple quantum circuits requiring very few classical computational resources. This makes it practical for use with large datasets.

\section{Results and discussion}
\label{Results}
In this section, we present the results of the classical CNN and the variations of the hybrid CNN. The performance of the models is evaluated against the core set of the 2020 PDBBind dataset. The training and validation steps are done with the refined set (separated into training and validation sets) for all the models. In order to reduce overfitting of the training data, we use an early stopping procedure, finishing the training step when the performance in the validation set has converged. To evaluate the convergence of the training process, we evaluate five error metrics:

\begin{itemize}
    \item \textbf{Root mean squared error (RMSE)}
    \item \textbf{Mean absolute error (MAE)}
    \item \textbf{Coefficient of determination R squared (R2)}: proportion of the variation of the dependent variable (binding affinity) that is predictable from the independent variable (prediction of the model).
    \item \textbf{Pearson correlation coefficient (Pearson)}: Linear correlation between two variables (binding affinity and prediction of the model). It ranges between $-1$ and $+1$.
    \item \textbf{Spearman coefficient}: Monotonic correlation coefficient. It ranges between $-1$ and $+1$. A Spearman correlation of $+1$ or $-1$ occurs when a variable is a perfect monotone function of the other.
\end{itemize}

\begin{figure*}
\includegraphics
   [width=0.80\textwidth]{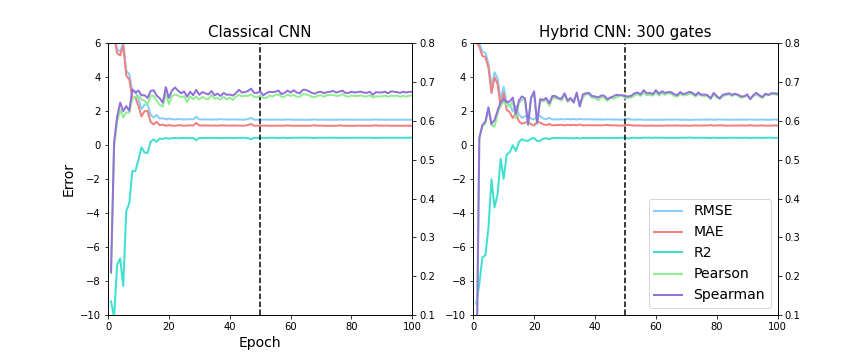}
\caption{Validation error metrics (RMSE, MAE, R2, Pearson and Spearman) evaluated in the validation set as a function of the training epochs. The left plot shows the training of the classical CNN and the right plot the hybrid CNN with 300 quantum gates. Both models converged after 50 epochs.}
\label{fig:validation}
\end{figure*}

Fig. \ref{fig:validation} shows the evolution of the five error metrics in the validation set, for the classical CNN and the hybrid CNN with 300 quantum gates. We see that for both cases, all the error metrics have stabilized after 50 epochs. Further training the models with the same data could lead to overfitting of the training data, decreasing the generalization capacity. For this reason, we stopped the training of all the models at 50 epochs. Fig. \ref{fig:validation} also shows that the Pearson and Spearman coefficients oscillate more than the other error metrics, even when the training has converged. For this reason, we believe that, in this case, the RMSE, MAE and R2 are better measurements of convergence of the models.

\begin{figure*}
\includegraphics
   [width=0.80\textwidth]{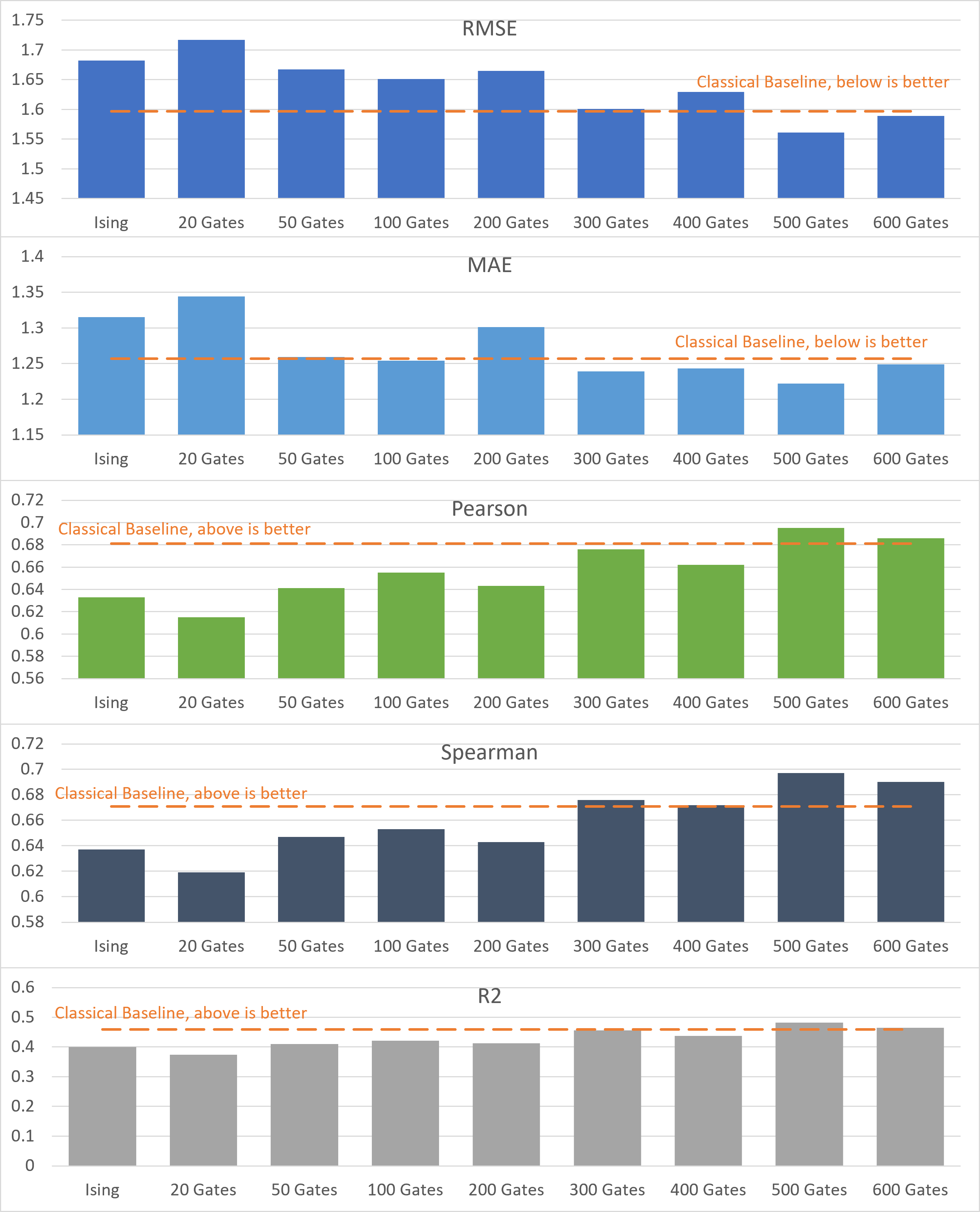}
\caption{Evaluation of the five error metrics (RMSE, MAE, R2, Pearson and Spearman) in the core set. Comparison of the hybrid CNN models constructed from the Ising model and the G3 family with 20,50,100,200,300,400,500 and 600 gates, together with the classical CNN. All the models have been trained with the refined set.}
\label{fig:test}
\end{figure*}

Once the models have been trained, we evaluate their performance on the test set. Fig. \ref{fig:test} shows the five error metrics, evaluated on the test set, for all the CNN models studied in this work. We compare the performance of the hybrid models with 20 - 600 quantum gates, with the performance of the classical CNN. The results show that, in general, the performance of the hybrid CNN models increases with the number of quantum gates until it reaches roughly the same performance as the classical CNN, at 300 quantum gates. From that point, the models with 400, 500 and 600 gates oscillate around the classical performance and do not improve significantly with the number of quantum gates. Therefore, we conclude that the number of quantum gates does affect the performance of the model, for shallow quantum circuits, and it stabilized when the quantum circuits achieve a certain depth. The minimal number of gates needed to achieve classical performance, in this case, would be around 300 quantum gates. Thus, for a certain choice of quantum circuits, decreasing the complexity of the CNN does not decrease its predictive performance. Fig. \ref{fig:test} also shows the performance of the hybrid model constructed from the Ising model. We see that the performance is a bit worse than the optimal G3 hybrid model.

The main motivation for designing a hybrid CNN model was to reduce the complexity and thus the training time of the neural network. A measure of complexity that does not depend on the hardware where the model is trained is the number of training parameters. Table \ref{tab:time} shows the training parameters of the classical CNN and all the hybrid CNN models. 

\begin{table}
    \centering
    \begin{tabular}{cccc}
    \hline \hline \\
        Hardware & Hybrid CNN & Classical CNN & Difference\\
        \hline\\
        Azure CPU & 10.7 days & 18.3days & 42\%\\
        Azure GPU & 24h & 39h & 38\%\\
        Purdue Anvil GPU & 16.3h & 22.1h & 26\%\\
        \hline \hline\\
        Training parameters & & & \\
        \hline\\
        All hardware & 8088499 & 10137129 & 20\%\\
         \hline
    \end{tabular}
    \caption{Training times and training parameters}
    \label{tab:time}
\end{table}

The classical CNN has around 10 million parameters, while the hybrid CNNs have around 8 million parameters, proving a 20\% reduction in model complexity. The training times of the models depend on the hardware where the training is executed. Training the CNNs with only CPUs requires much longer execution times. Using GPUs highly accelerates the training process, reducing the training time from many days to hours. In our experiments, the models have been trained using only CPUs and with two types of GPUs. The details of the used hardware is shown in Table \ref{tab:hardware}. 

\begin{table*}
    \centering
    \begin{tabular}{ m{1.5cm}|m{4.5cm}|m{4cm}|m{4cm}} 
    \hline \hline \\
         & Azure CPU & Azure GPU & Purdue Anvil GPU\\
        \hline\\
        CPU &	Intel Xeon E5-2673 v3 2.4 GHz &	Intel Xeon E5-2690 v3 & 2 x 3rd Gen AMD EPYC 7763\\
        Cores	& 4	& 6	& 128\\
        GPU	& -	& 1 x NVIDIA Tesla K80	& 4 x NVIDIA A100\\
        Memory	& 14GB	& 56GB	& 512GB\\
         \hline
    \end{tabular}
    \caption{Hardware specification}
    \label{tab:hardware}
\end{table*}

The improvement in training times with the hybrid model over the classical varies from 26\% to 42\%. Using more powerful GPUs reduces the difference in training times, but the hardware is also more costly. The difference in training times in all cases is limited by the difference in training parameters, which is a hardware-agnostic measure of complexity. 

\begin{figure*}
\includegraphics
   [width=0.80\textwidth]{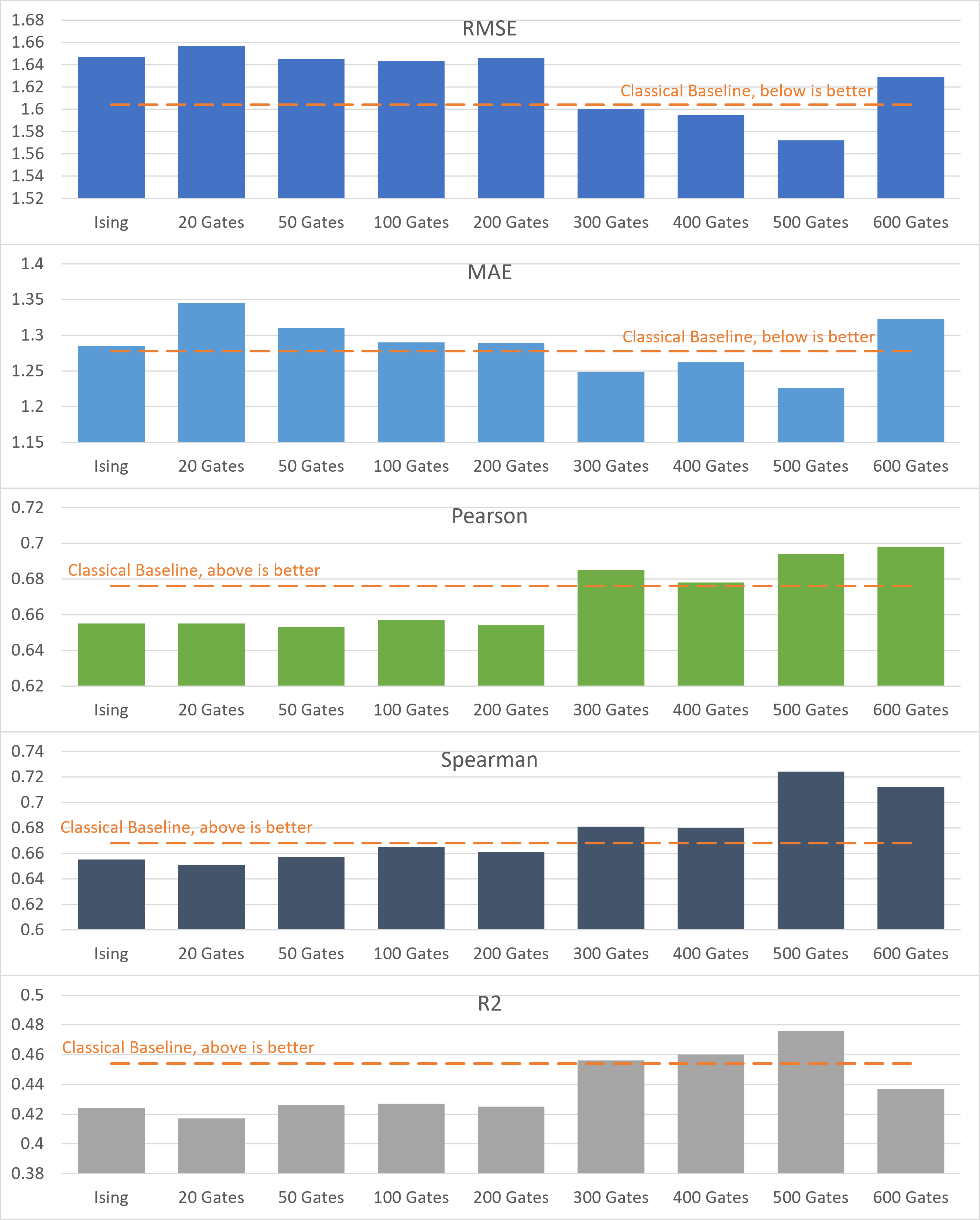}
\caption{Evaluation of the five error metrics (RMSE, MAE, R2, Pearson and Spearman) in the core set. Comparison of the hybrid CNN models constructed from the Ising model and the G3 family with 20,50,100,200,300,400,500 and 600 gates, together with the classical CNN. All the models have been trained with the general set.}
\label{fig:test_general}
\end{figure*}

After analyzing the results from the models trained with the refined set, we repeated the experiments training the models with the general set. The general set has almost three times more data than the refined set, and thus the training takes more time and computational resources. We observed that the models required more epochs for the performance to converge in the validation set. The performance metrics evaluated in the test set are displayed in Fig. \ref{fig:test_general}. We see that the performance results are equivalent to the ones from the models trained with the refined set. The performance of the hybrid G3 models increases with the number of quantum gates until it converges at around 300 gates. Then, the performance oscillates around the classical performance. The Ising model has suboptimal performance compared to the classical CNN or the hybrid G3 CNN with 300 gates. We conclude that training the models with the general set leads to equivalent results to training the models with the refined set, but it requires longer training times and having more computational resources. \\
\\
CNNs are widely used models to learn from data such as time series, images or volumetric representations. Their goal is to unravel hidden patterns from the input data and use them to predict the target. Thus, the complexity of a CNN model highly depends on the complexity of the data. Hybrid quantum-classical CNN models can help reduce the number of parameters of the neural network while maintaining its prediction capacity. One natural question would be how this reduction of training parameters scales with the size of the data. Let's consider that each sample has size $(C,N,N,N)$, where $C$ is the number of features and $N$ is the size of the volume side. The reduction of model complexity corresponds to the number of parameters of the first layer of the network. Therefore, the reduction of training parameters scales linearly with the number of features $C$. The number of training parameters does not explicitly depend on $N$, because each filter is applied locally to a portion of the data, as many times as needed to cover the whole sample. However, when the dimensionality of the data increases, usually more filters are needed for the CNN to converge. As the data complexity increases, more complex models are needed to learn useful information from it.

\begin{figure*}
    \centering
    \includegraphics[width=0.95\textwidth]{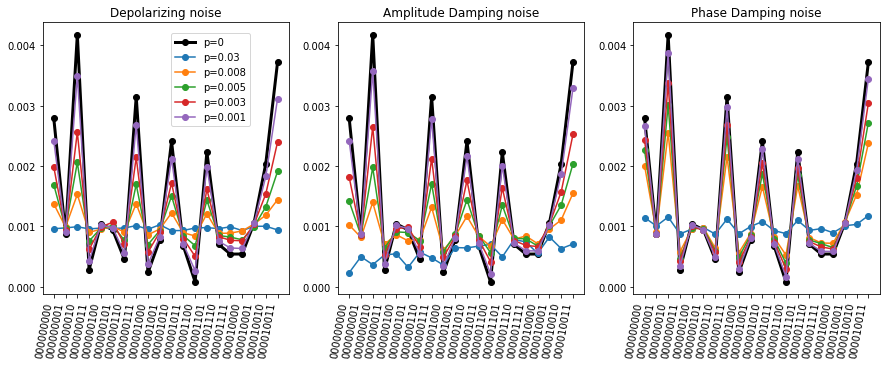}
    \caption{Example of an output of the quantum circuit used in the quantum convolutional layer for three noise models and different error rates. For easier visualization, only the first 20 outputs are displayed in the figure.}
    \label{fig:counts_noise}
\end{figure*}

\begin{figure*}
    \centering
    \includegraphics[width=0.95\textwidth]{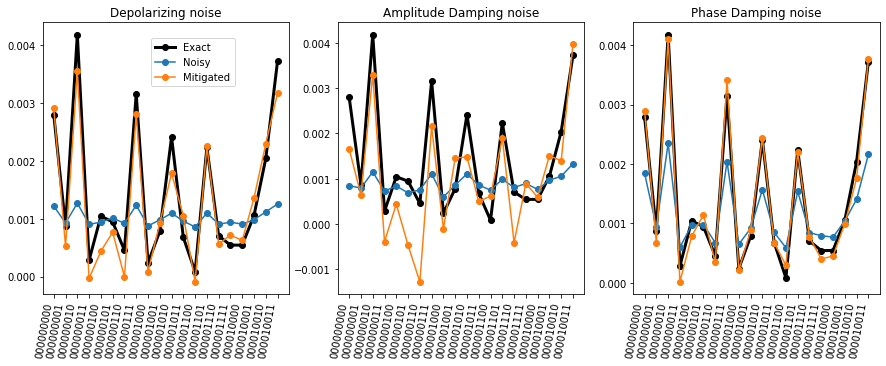}
    \caption{Example of an output of the quantum circuit used in the quantum convolutional layer for three noise models and error rate $p=0.01$, together with the output of the error mitigation algorithm. For easier visualization, only the first 20 outputs are displayed in the figure.}
    \label{fig:counts_mitig}
\end{figure*}

\begin{table*}
    \centering
    \begin{tabular}{|c|c|c|c|c|c|}
    \hline
    \hline
         Error model & Error rate $p$ & Regularization $\alpha$ & MSE Noisy circuits & MSE Mitigated circuits & Tendency accuracy  \\
         \hline
         \multirow{6}{*}{Depolarizing} & 0.03 & 0.1 & $9.7 \cdot 10^{-7}$ &  $9.7 \cdot 10^{-7}$ & 0.59\\
          & 0.01 & 0.01 & $7.7 \cdot 10^{-7}$ &  $2.1 \cdot 10^{-7}$ & 0.75\\
          & 0.008 & $1 \cdot 10^{-5}$ & $6.7 \cdot 10^{-7}$ &  $1.0 \cdot 10^{-7}$ & 0.80\\
          & 0.005 & $1 \cdot 10^{-5}$ & $4.3 \cdot 10^{-7}$ &  $3.4 \cdot 10^{-8}$ & 0.84\\
          & 0.003 & $1 \cdot 10^{-5}$ & $2.3 \cdot 10^{-7}$ &  $1.1 \cdot 10^{-8}$ & 0.87\\
          & 0.001 & $1 \cdot 10^{-6}$ & $4.0 \cdot 10^{-8}$ &  $1.4 \cdot 10^{-9}$ & 0.89\\
         \hline
         \multirow{6}{*}{Amplitude Damping} & 0.03 & 0.1 & $1.1 \cdot 10^{-6}$ &  $9.9 \cdot 10^{-7}$ & 0.54\\
          & 0.01 & $1 \cdot 10^{-4}$ & $7.3 \cdot 10^{-7}$ &  $5.3 \cdot 10^{-7}$ & 0.62\\
          & 0.008 & $1 \cdot 10^{-4}$ & $6.2 \cdot 10^{-7}$ &  $3.4 \cdot 10^{-7}$ & 0.66\\
          & 0.005 & $1 \cdot 10^{-5}$ & $3.9 \cdot 10^{-7}$ &  $6.8 \cdot 10^{-8}$ & 0.75\\
          & 0.003 & $1 \cdot 10^{-5}$ & $2.0 \cdot 10^{-7}$ &  $2.1 \cdot 10^{-8}$ & 0.80\\
          & 0.001 & $1 \cdot 10^{-5}$ & $3.2 \cdot 10^{-8}$ &  $2.4 \cdot 10^{-9}$ & 0.84\\
         \hline
         \multirow{6}{*}{Phase Damping} & 0.03 & $1 \cdot 10^{-5}$ & $8.2 \cdot 10^{-7}$ &  $4.0 \cdot 10^{-7}$ & 0.67\\
          & 0.01 & $1 \cdot 10^{-5}$ & $3.2 \cdot 10^{-7}$ &  $3.7 \cdot 10^{-8}$ & 0.81\\
          & 0.008 & $1 \cdot 10^{-5}$ & $2.4 \cdot 10^{-7}$ &  $2.4 \cdot 10^{-8}$ & 0.82\\
          & 0.005 & $1 \cdot 10^{-5}$ & $1.2 \cdot 10^{-7}$ &  $9.4 \cdot 10^{-9}$ & 0.85\\
          & 0.003 & $1 \cdot 10^{-5}$ & $5.1 \cdot 10^{-8}$ &  $3.6 \cdot 10^{-9}$ & 0.85\\
          & 0.001 & $1 \cdot 10^{-6}$ & $6.9 \cdot 10^{-9}$ &  $5.0 \cdot 10^{-10}$ & 0.85\\
         \hline
    \end{tabular}
    \caption{Performance of the error mitigation algorithm by means of the mean squared error (MSE) and tendency accuracy for different noise models and error rates.}
    \label{tab:error_mitig}
\end{table*}

After analyzing the results of the noiseless quantum circuits, we perform noisy simulations for three different quantum channels and analyze the corresponding performance. An example of the output of a quantum circuit for different noise models and different error probabilities is shown in Fig. \ref{fig:counts_noise}. We see that the three noise models reduce the probability amplitude of the circuits' outputs. The main difference between the behavior of the noise models is that the phase damping channel reduces the probability amplitude slower than the other two models.  In all cases, when the error probability reaches $p=0.03$, the quantum information is lost, since the amplitude peaks can no longer be distinguished. On the other hand, when the error rate is smaller than $p=0.03$, the DRER algorithm can successfully mitigate the noisy outputs. An example of the performance of the DRER algorithm for $p=0.01$ is shown in Fig. \ref{fig:counts_mitig}. Even though the noise of the quantum device significantly reduces the amplitudes of the distribution, the DRER algorithm can recover the original amplitudes with significant accuracy. For every noise model and error rate $p$, we performed a hryperparameter optimization to obtain the best linear model to mitigate the quantum errors. The results are shown in Table \ref{tab:error_mitig}. Apart from evaluating the mean squared error (MSE), we also evaluate the tendency accuracy, that is, the proportion of times the DRER algorithm modifies the output in the correct direction. Let $y_\text{noisy}$, $y_\text{noiseless}$, $y_\text{mitigated}$ be the noisy, noiseless and mitigated counts respectively. Then, the tendency accuracy measures the proportion of times $|y_\text{mitigated} - y_\text{noiseless}| < |y_\text{noisy} - y_\text{noiseless}|$. Table \ref{tab:error_mitig} shows that when $p \leq 0.01$ the MSE of the mitigated circuits is smaller than the MSE of the noisy circuits. On the other hand, for $p=0.03$ the MSE of the mitigated circuits is similar or even larger than the MSE of the noisy circuits, and the tendency accuracy is barely better than random guessing. This result agrees with Fig. \ref{fig:counts_noise}, since the noisy simulations with $p=0.03$ are basically a constant value. Table \ref{tab:error_mitig} also shows that the tendency accuracy increases when the error probability $p$ decreases. For the depolarizing quantum channel, the tendency accuracy reaches the value $0.8$ when $p=0.008$, and increases to $0.89$ with $p=0.001$. The amplitude damping noise seems to be the hardest to mitigate, since the tendency accuracy increases slower than the tendency accuracy of the other noise models. This is due to the fact that the amplitude damping channel introduces non-zero counts apart from mitigating the amplitudes of the noiseless simulation. On the other hand, the tendency accuracy increases faster with the phase damping channel, where it reaches the value $0.81$ with $p=0.01$. All in all, these results show that as long as the error rates are smaller than $p=0.01$, the DRER algorithm can successfully mitigate the errors introduced by the quantum device on the quantum convolutional layer with 300 gates.

\section{Conclusions}
\label{Conclusions}

Understanding the binding affinity of a drug candidate can provide valuable insights into its potential efficacy and help to identify potential side effects. Additionally, predicting the binding affinity can help with the design of new molecules that bind more strongly to their target protein. This is especially important in the drug development process when dealing with new treatments that work on previously unexplored biological mechanisms. For this reason, designing efficient computational methods that can accurately predict the binding affinity between a molecule and a target protein is essential to speed up the drug discovery process. Deep learning methods such as 3D CNNs provide promising results in this aspect, since they learn directly from the atomic structure of the protein-ligand pair. However, one of the biggest challenges of deep learning methods is the high complexity of the networks, which require learning millions of training parameters. This fact makes the training process long and costly, limiting the exploration of different network architectures. Quantum machine learning is a field that seeks to leverage the advantages of quantum computing to improve machine learning algorithms. Because of the exponential scaling of the Hilbert space, quantum computers have the ability to handle large and high-dimensional datasets and speed up machine learning algorithms. In this work we present a hybrid quantum-classical 3D CNN, which reduces the complexity of the classical 3D CNN while maintaining an optimal prediction performance. With proper design of quantum circuits, the hybrid CNN reduces the number of training parameters by 20\%, which implies a reduction of training times of 20\%-40\%, depending on the hardware where the algorithm is executed. Apart from testing the performance of the algorithm in classical hardware, our work also proves the potential effectiveness of the method with noisy real hardware aided with a simple error mitigation technique. Our results show that if the error probability is smaller than 0.01, a commonly-used error mitigation technique can accurately recover the noiseless outputs of the quantum circuit. All in all, this work shows how quantum machine learning offers the potential to reduce the complexity and long training times of classical neural networks by leveraging the advantages of quantum computing to handle large and high-dimensional datasets and speed up machine learning algorithms.

\begin{acknowledgments}
The project that gave rise to these results received the support of a fellowship from la Caixa Foundation (ID 100010434). The fellowship code is LCF/BQ/DR20/11790028.

\end{acknowledgments}

\bibliography{aipsamp}

\begin{thebibliography}{32}%
\makeatletter
\providecommand \@ifxundefined [1]{%
 \@ifx{#1\undefined}
}%
\providecommand \@ifnum [1]{%
 \ifnum #1\expandafter \@firstoftwo
 \else \expandafter \@secondoftwo
 \fi
}%
\providecommand \@ifx [1]{%
 \ifx #1\expandafter \@firstoftwo
 \else \expandafter \@secondoftwo
 \fi
}%
\providecommand \natexlab [1]{#1}%
\providecommand \enquote  [1]{``#1''}%
\providecommand \bibnamefont  [1]{#1}%
\providecommand \bibfnamefont [1]{#1}%
\providecommand \citenamefont [1]{#1}%
\providecommand \href@noop [0]{\@secondoftwo}%
\providecommand \href [0]{\begingroup \@sanitize@url \@href}%
\providecommand \@href[1]{\@@startlink{#1}\@@href}%
\providecommand \@@href[1]{\endgroup#1\@@endlink}%
\providecommand \@sanitize@url [0]{\catcode `\\12\catcode `\$12\catcode
  `\&12\catcode `\#12\catcode `\^12\catcode `\_12\catcode `\%12\relax}%
\providecommand \@@startlink[1]{}%
\providecommand \@@endlink[0]{}%
\providecommand \url  [0]{\begingroup\@sanitize@url \@url }%
\providecommand \@url [1]{\endgroup\@href {#1}{\urlprefix }}%
\providecommand \urlprefix  [0]{URL }%
\providecommand \Eprint [0]{\href }%
\providecommand \doibase [0]{http://dx.doi.org/}%
\providecommand \selectlanguage [0]{\@gobble}%
\providecommand \bibinfo  [0]{\@secondoftwo}%
\providecommand \bibfield  [0]{\@secondoftwo}%
\providecommand \translation [1]{[#1]}%
\providecommand \BibitemOpen [0]{}%
\providecommand \bibitemStop [0]{}%
\providecommand \bibitemNoStop [0]{.\EOS\space}%
\providecommand \EOS [0]{\spacefactor3000\relax}%
\providecommand \BibitemShut  [1]{\csname bibitem#1\endcsname}%
\let\auto@bib@innerbib\@empty
\bibitem [{\citenamefont {Jones}\ \emph {et~al.}(2021)\citenamefont {Jones},
  \citenamefont {Kim}, \citenamefont {Zhang}, \citenamefont {Zemla},
  \citenamefont {Stevenson}, \citenamefont {Bennett}, \citenamefont {Kirshner},
  \citenamefont {Wong}, \citenamefont {Lightstone},\ and\ \citenamefont
  {Allen}}]{ATOM}%
  \BibitemOpen
  \bibfield  {author} {\bibinfo {author} {\bibfnamefont {D.}~\bibnamefont
  {Jones}}, \bibinfo {author} {\bibfnamefont {H.}~\bibnamefont {Kim}}, \bibinfo
  {author} {\bibfnamefont {X.}~\bibnamefont {Zhang}}, \bibinfo {author}
  {\bibfnamefont {A.}~\bibnamefont {Zemla}}, \bibinfo {author} {\bibfnamefont
  {G.}~\bibnamefont {Stevenson}}, \bibinfo {author} {\bibfnamefont {W.~F.~D.}\
  \bibnamefont {Bennett}}, \bibinfo {author} {\bibfnamefont {D.}~\bibnamefont
  {Kirshner}}, \bibinfo {author} {\bibfnamefont {S.~E.}\ \bibnamefont {Wong}},
  \bibinfo {author} {\bibfnamefont {F.~C.}\ \bibnamefont {Lightstone}}, \ and\
  \bibinfo {author} {\bibfnamefont {J.~E.}\ \bibnamefont {Allen}},\ }\href
  {\doibase 10.1021/acs.jcim.0c01306} {\bibfield  {journal} {\bibinfo
  {journal} {Journal of Chemical Information and Modeling}\ }\textbf {\bibinfo
  {volume} {61}},\ \bibinfo {pages} {1583} (\bibinfo {year}
  {2021})}\BibitemShut {NoStop}%
\bibitem [{\citenamefont {Wallach}\ \emph {et~al.}(2015)\citenamefont
  {Wallach}, \citenamefont {Dzamba},\ and\ \citenamefont {Heifets}}]{AtomNet}%
  \BibitemOpen
  \bibfield  {author} {\bibinfo {author} {\bibfnamefont {I.}~\bibnamefont
  {Wallach}}, \bibinfo {author} {\bibfnamefont {M.}~\bibnamefont {Dzamba}}, \
  and\ \bibinfo {author} {\bibfnamefont {A.}~\bibnamefont {Heifets}},\
  }\href@noop {} {\bibfield  {journal} {\bibinfo  {journal} {ArXiv}\ }\textbf
  {\bibinfo {volume} {abs/1510.02855}} (\bibinfo {year} {2015})}\BibitemShut
  {NoStop}%
\bibitem [{\citenamefont {Kuzminykh}\ \emph {et~al.}(2018)\citenamefont
  {Kuzminykh}, \citenamefont {Polykovskiy}, \citenamefont {Kadurin},
  \citenamefont {Zhebrak}, \citenamefont {Baskov}, \citenamefont {Nikolenko},
  \citenamefont {Shayakhmetov},\ and\ \citenamefont {Zhavoronkov}}]{Wave}%
  \BibitemOpen
  \bibfield  {author} {\bibinfo {author} {\bibfnamefont {D.}~\bibnamefont
  {Kuzminykh}}, \bibinfo {author} {\bibfnamefont {D.}~\bibnamefont
  {Polykovskiy}}, \bibinfo {author} {\bibfnamefont {A.}~\bibnamefont
  {Kadurin}}, \bibinfo {author} {\bibfnamefont {A.}~\bibnamefont {Zhebrak}},
  \bibinfo {author} {\bibfnamefont {I.}~\bibnamefont {Baskov}}, \bibinfo
  {author} {\bibfnamefont {S.}~\bibnamefont {Nikolenko}}, \bibinfo {author}
  {\bibfnamefont {R.}~\bibnamefont {Shayakhmetov}}, \ and\ \bibinfo {author}
  {\bibfnamefont {A.}~\bibnamefont {Zhavoronkov}},\ }\href {\doibase
  10.1021/acs.molpharmaceut.7b01134} {\bibfield  {journal} {\bibinfo  {journal}
  {Molecular Pharmaceutics}\ }\textbf {\bibinfo {volume} {15}},\ \bibinfo
  {pages} {4378} (\bibinfo {year} {2018})}\BibitemShut {NoStop}%
\bibitem [{\citenamefont {Jiménez}\ \emph {et~al.}(2017)\citenamefont
  {Jiménez}, \citenamefont {Doerr}, \citenamefont {Martínez-Rosell},
  \citenamefont {Rose},\ and\ \citenamefont {De~Fabritiis}}]{DeepSite}%
  \BibitemOpen
  \bibfield  {author} {\bibinfo {author} {\bibfnamefont {J.}~\bibnamefont
  {Jiménez}}, \bibinfo {author} {\bibfnamefont {S.}~\bibnamefont {Doerr}},
  \bibinfo {author} {\bibfnamefont {G.}~\bibnamefont {Martínez-Rosell}},
  \bibinfo {author} {\bibfnamefont {A.~S.}\ \bibnamefont {Rose}}, \ and\
  \bibinfo {author} {\bibfnamefont {G.}~\bibnamefont {De~Fabritiis}},\ }\href
  {\doibase 10.1093/bioinformatics/btx350} {\bibfield  {journal} {\bibinfo
  {journal} {Bioinformatics}\ }\textbf {\bibinfo {volume} {33}},\ \bibinfo
  {pages} {3036} (\bibinfo {year} {2017})}\BibitemShut {NoStop}%
\bibitem [{\citenamefont {Ragoza}\ \emph {et~al.}(2017)\citenamefont {Ragoza},
  \citenamefont {Hochuli}, \citenamefont {Idrobo}, \citenamefont {Sunseri},\
  and\ \citenamefont {Koes}}]{ScoringCNN}%
  \BibitemOpen
  \bibfield  {author} {\bibinfo {author} {\bibfnamefont {M.}~\bibnamefont
  {Ragoza}}, \bibinfo {author} {\bibfnamefont {J.}~\bibnamefont {Hochuli}},
  \bibinfo {author} {\bibfnamefont {E.}~\bibnamefont {Idrobo}}, \bibinfo
  {author} {\bibfnamefont {J.}~\bibnamefont {Sunseri}}, \ and\ \bibinfo
  {author} {\bibfnamefont {D.~R.}\ \bibnamefont {Koes}},\ }\href {\doibase
  10.1021/acs.jcim.6b00740} {\bibfield  {journal} {\bibinfo  {journal} {Journal
  of Chemical Information and Modeling}\ }\textbf {\bibinfo {volume} {57}},\
  \bibinfo {pages} {942} (\bibinfo {year} {2017})}\BibitemShut {NoStop}%
\bibitem [{\citenamefont {Jiménez}\ \emph
  {et~al.}(2018{\natexlab{a}})\citenamefont {Jiménez}, \citenamefont
  {Škalič}, \citenamefont {Martínez-Rosell},\ and\ \citenamefont
  {De~Fabritiis}}]{AbsoluteCNN}%
  \BibitemOpen
  \bibfield  {author} {\bibinfo {author} {\bibfnamefont {J.}~\bibnamefont
  {Jiménez}}, \bibinfo {author} {\bibfnamefont {M.}~\bibnamefont {Škalič}},
  \bibinfo {author} {\bibfnamefont {G.}~\bibnamefont {Martínez-Rosell}}, \
  and\ \bibinfo {author} {\bibfnamefont {G.}~\bibnamefont {De~Fabritiis}},\
  }\href {\doibase 10.1021/acs.jcim.7b00650} {\bibfield  {journal} {\bibinfo
  {journal} {Journal of Chemical Information and Modeling}\ }\textbf {\bibinfo
  {volume} {58}},\ \bibinfo {pages} {287} (\bibinfo {year}
  {2018}{\natexlab{a}})}\BibitemShut {NoStop}%
\bibitem [{\citenamefont {Mohri}\ \emph {et~al.}(2018)\citenamefont {Mohri},
  \citenamefont {Rostamizadeh},\ and\ \citenamefont {Talwalkar}}]{MLBook}%
  \BibitemOpen
  \bibfield  {author} {\bibinfo {author} {\bibfnamefont {M.}~\bibnamefont
  {Mohri}}, \bibinfo {author} {\bibfnamefont {A.}~\bibnamefont {Rostamizadeh}},
  \ and\ \bibinfo {author} {\bibfnamefont {A.}~\bibnamefont {Talwalkar}},\
  }\href@noop {} {\emph {\bibinfo {title} {Foundations of Machine Learning}}},\
  \bibinfo {edition} {2nd}\ ed.\ (\bibinfo  {publisher} {The MIT Press},\
  \bibinfo {year} {2018})\BibitemShut {NoStop}%
\bibitem [{\citenamefont {Czarnik}\ \emph {et~al.}(2021)\citenamefont
  {Czarnik}, \citenamefont {Arrasmith}, \citenamefont {Coles},\ and\
  \citenamefont {Cincio}}]{DRER}%
  \BibitemOpen
  \bibfield  {author} {\bibinfo {author} {\bibfnamefont {P.}~\bibnamefont
  {Czarnik}}, \bibinfo {author} {\bibfnamefont {A.}~\bibnamefont {Arrasmith}},
  \bibinfo {author} {\bibfnamefont {P.~J.}\ \bibnamefont {Coles}}, \ and\
  \bibinfo {author} {\bibfnamefont {L.}~\bibnamefont {Cincio}},\ }\href
  {\doibase 10.22331/q-2021-11-26-592} {\bibfield  {journal} {\bibinfo
  {journal} {Quantum}\ }\textbf {\bibinfo {volume} {5}},\ \bibinfo {pages}
  {592} (\bibinfo {year} {2021})}\BibitemShut {NoStop}%
\bibitem [{\citenamefont {wwPDB consortium}(2018)}]{PDBBind}%
  \BibitemOpen
  \bibfield  {author} {\bibinfo {author} {\bibnamefont {wwPDB consortium}},\
  }\href {\doibase 10.1093/nar/gky949} {\bibfield  {journal} {\bibinfo
  {journal} {Nucleic Acids Research}\ }\textbf {\bibinfo {volume} {47}},\
  \bibinfo {pages} {D520} (\bibinfo {year} {2018})}\BibitemShut {NoStop}%
\bibitem [{\citenamefont {Jiménez}\ \emph
  {et~al.}(2018{\natexlab{b}})\citenamefont {Jiménez}, \citenamefont
  {Škalič}, \citenamefont {Martínez-Rosell},\ and\ \citenamefont
  {De~Fabritiis}}]{3DCNNBA}%
  \BibitemOpen
  \bibfield  {author} {\bibinfo {author} {\bibfnamefont {J.}~\bibnamefont
  {Jiménez}}, \bibinfo {author} {\bibfnamefont {M.}~\bibnamefont {Škalič}},
  \bibinfo {author} {\bibfnamefont {G.}~\bibnamefont {Martínez-Rosell}}, \
  and\ \bibinfo {author} {\bibfnamefont {G.}~\bibnamefont {De~Fabritiis}},\
  }\href {\doibase 10.1021/acs.jcim.7b00650} {\bibfield  {journal} {\bibinfo
  {journal} {Journal of Chemical Information and Modeling}\ }\textbf {\bibinfo
  {volume} {58}},\ \bibinfo {pages} {287} (\bibinfo {year}
  {2018}{\natexlab{b}})},\ \bibinfo {note} {pMID: 29309725}\BibitemShut
  {NoStop}%
\bibitem [{\citenamefont {Zhang}\ \emph {et~al.}(2019)\citenamefont {Zhang},
  \citenamefont {Liao}, \citenamefont {KM}, \citenamefont {Yin},\ and\
  \citenamefont {Wei}}]{DeepLearningBA}%
  \BibitemOpen
  \bibfield  {author} {\bibinfo {author} {\bibfnamefont {H.}~\bibnamefont
  {Zhang}}, \bibinfo {author} {\bibfnamefont {L.}~\bibnamefont {Liao}},
  \bibinfo {author} {\bibfnamefont {S.}~\bibnamefont {KM}}, \bibinfo {author}
  {\bibfnamefont {P.}~\bibnamefont {Yin}}, \ and\ \bibinfo {author}
  {\bibfnamefont {Y.}~\bibnamefont {Wei}},\ }\href {\doibase
  10.7717/peerj.7362} {\bibfield  {journal} {\bibinfo  {journal} {PeerJ}\
  }\textbf {\bibinfo {volume} {7}} (\bibinfo {year} {2019}),\
  10.7717/peerj.7362}\BibitemShut {NoStop}%
\bibitem [{\citenamefont {Wang}\ \emph {et~al.}(2021)\citenamefont {Wang},
  \citenamefont {Chan},\ and\ \citenamefont {Yan}}]{Fingerprints}%
  \BibitemOpen
  \bibfield  {author} {\bibinfo {author} {\bibfnamefont {D.~D.}\ \bibnamefont
  {Wang}}, \bibinfo {author} {\bibfnamefont {M.-T.}\ \bibnamefont {Chan}}, \
  and\ \bibinfo {author} {\bibfnamefont {H.}~\bibnamefont {Yan}},\ }\href
  {\doibase https://doi.org/10.1016/j.csbj.2021.11.018} {\bibfield  {journal}
  {\bibinfo  {journal} {Computational and Structural Biotechnology Journal}\
  }\textbf {\bibinfo {volume} {19}},\ \bibinfo {pages} {6291} (\bibinfo {year}
  {2021})}\BibitemShut {NoStop}%
\bibitem [{\citenamefont {Stepniewska-Dziubinska}\ \emph
  {et~al.}(2018)\citenamefont {Stepniewska-Dziubinska}, \citenamefont
  {Zielenkiewicz},\ and\ \citenamefont {Siedlecki}}]{Pafnucy}%
  \BibitemOpen
  \bibfield  {author} {\bibinfo {author} {\bibfnamefont {M.~M.}\ \bibnamefont
  {Stepniewska-Dziubinska}}, \bibinfo {author} {\bibfnamefont {P.}~\bibnamefont
  {Zielenkiewicz}}, \ and\ \bibinfo {author} {\bibfnamefont {P.}~\bibnamefont
  {Siedlecki}},\ }\href {\doibase 10.1093/bioinformatics/bty374} {\bibfield
  {journal} {\bibinfo  {journal} {Bioinformatics}\ }\textbf {\bibinfo {volume}
  {34}},\ \bibinfo {pages} {3666} (\bibinfo {year} {2018})}\BibitemShut
  {NoStop}%
\bibitem [{\citenamefont {Lu}\ \emph {et~al.}(2019)\citenamefont {Lu},
  \citenamefont {Wang}, \citenamefont {Zhang}, \citenamefont {Yoon},\ and\
  \citenamefont {Won}}]{ImageSegm}%
  \BibitemOpen
  \bibfield  {author} {\bibinfo {author} {\bibfnamefont {H.}~\bibnamefont
  {Lu}}, \bibinfo {author} {\bibfnamefont {H.}~\bibnamefont {Wang}}, \bibinfo
  {author} {\bibfnamefont {Q.}~\bibnamefont {Zhang}}, \bibinfo {author}
  {\bibfnamefont {S.~W.}\ \bibnamefont {Yoon}}, \ and\ \bibinfo {author}
  {\bibfnamefont {D.}~\bibnamefont {Won}},\ }\href {\doibase
  https://doi.org/10.1016/j.promfg.2020.01.386} {\bibfield  {journal} {\bibinfo
   {journal} {Procedia Manufacturing}\ }\textbf {\bibinfo {volume} {39}},\
  \bibinfo {pages} {422} (\bibinfo {year} {2019})}\BibitemShut {NoStop}%
\bibitem [{\citenamefont {Singh}(2020)}]{medicalImages}%
  \BibitemOpen
  \bibfield  {author} {\bibinfo {author} {\bibfnamefont {S.}~\bibnamefont
  {Singh}},\ }\href {\doibase 10.3390/s20185097} {\bibfield  {journal}
  {\bibinfo  {journal} {Sensors}\ }\textbf {\bibinfo {volume} {20}} (\bibinfo
  {year} {2020}),\ 10.3390/s20185097}\BibitemShut {NoStop}%
\bibitem [{\citenamefont {Arunnehru}\ \emph {et~al.}(2018)\citenamefont
  {Arunnehru}, \citenamefont {Chamundeeswari},\ and\ \citenamefont
  {Bharathi}}]{HumanAction}%
  \BibitemOpen
  \bibfield  {author} {\bibinfo {author} {\bibfnamefont {J.}~\bibnamefont
  {Arunnehru}}, \bibinfo {author} {\bibfnamefont {G.}~\bibnamefont
  {Chamundeeswari}}, \ and\ \bibinfo {author} {\bibfnamefont {S.~P.}\
  \bibnamefont {Bharathi}},\ }\href {\doibase
  https://doi.org/10.1016/j.procs.2018.07.059} {\bibfield  {journal} {\bibinfo
  {journal} {Procedia Computer Science}\ }\textbf {\bibinfo {volume} {133}},\
  \bibinfo {pages} {471} (\bibinfo {year} {2018})},\ \bibinfo {note}
  {international Conference on Robotics and Smart Manufacturing
  (RoSMa2018)}\BibitemShut {NoStop}%
\bibitem [{\citenamefont {He}\ \emph {et~al.}(2016)\citenamefont {He},
  \citenamefont {Zhang}, \citenamefont {Ren},\ and\ \citenamefont
  {Sun}}]{ResNet}%
  \BibitemOpen
  \bibfield  {author} {\bibinfo {author} {\bibfnamefont {K.}~\bibnamefont
  {He}}, \bibinfo {author} {\bibfnamefont {X.}~\bibnamefont {Zhang}}, \bibinfo
  {author} {\bibfnamefont {S.}~\bibnamefont {Ren}}, \ and\ \bibinfo {author}
  {\bibfnamefont {J.}~\bibnamefont {Sun}},\ }in\ \href {\doibase
  10.1109/CVPR.2016.90} {\emph {\bibinfo {booktitle} {2016 IEEE Conference on
  Computer Vision and Pattern Recognition (CVPR)}}}\ (\bibinfo {year} {2016})\
  pp.\ \bibinfo {pages} {770--778}\BibitemShut {NoStop}%
\bibitem [{\citenamefont {Oh}\ \emph {et~al.}(2020)\citenamefont {Oh},
  \citenamefont {Choi},\ and\ \citenamefont {Kim}}]{tutorialQCNN}%
  \BibitemOpen
  \bibfield  {author} {\bibinfo {author} {\bibfnamefont {S.}~\bibnamefont
  {Oh}}, \bibinfo {author} {\bibfnamefont {J.}~\bibnamefont {Choi}}, \ and\
  \bibinfo {author} {\bibfnamefont {J.}~\bibnamefont {Kim}}\ }(\bibinfo {year}
  {2020})\ pp.\ \bibinfo {pages} {236--239}\BibitemShut {NoStop}%
\bibitem [{\citenamefont {Chen}\ \emph {et~al.}(2022)\citenamefont {Chen},
  \citenamefont {Wei}, \citenamefont {Zhang}, \citenamefont {Yu},\ and\
  \citenamefont {Yoo}}]{QCNN}%
  \BibitemOpen
  \bibfield  {author} {\bibinfo {author} {\bibfnamefont {S.~Y.-C.}\
  \bibnamefont {Chen}}, \bibinfo {author} {\bibfnamefont {T.-C.}\ \bibnamefont
  {Wei}}, \bibinfo {author} {\bibfnamefont {C.}~\bibnamefont {Zhang}}, \bibinfo
  {author} {\bibfnamefont {H.}~\bibnamefont {Yu}}, \ and\ \bibinfo {author}
  {\bibfnamefont {S.}~\bibnamefont {Yoo}},\ }\href {\doibase
  10.1103/PhysRevResearch.4.013231} {\bibfield  {journal} {\bibinfo  {journal}
  {Phys. Rev. Research}\ }\textbf {\bibinfo {volume} {4}},\ \bibinfo {pages}
  {013231} (\bibinfo {year} {2022})}\BibitemShut {NoStop}%
\bibitem [{\citenamefont {Henderson}\ \emph {et~al.}(2020)\citenamefont
  {Henderson}, \citenamefont {Shakya}, \citenamefont {Pradhan},\ and\
  \citenamefont {Cook}}]{Quanvolutional}%
  \BibitemOpen
  \bibfield  {author} {\bibinfo {author} {\bibfnamefont {M.}~\bibnamefont
  {Henderson}}, \bibinfo {author} {\bibfnamefont {S.}~\bibnamefont {Shakya}},
  \bibinfo {author} {\bibfnamefont {S.}~\bibnamefont {Pradhan}}, \ and\
  \bibinfo {author} {\bibfnamefont {T.}~\bibnamefont {Cook}},\ }\href {\doibase
  10.1007/s42484-020-00012-y} {\bibfield  {journal} {\bibinfo  {journal}
  {Quantum Machine Intelligence}\ }\textbf {\bibinfo {volume} {2}},\ \bibinfo
  {pages} {1} (\bibinfo {year} {2020})}\BibitemShut {NoStop}%
\bibitem [{\citenamefont {Araujo}\ \emph {et~al.}(2021)\citenamefont {Araujo},
  \citenamefont {Park}, \citenamefont {Petruccione},\ and\ \citenamefont
  {da~Silva}}]{amplitudeEncoding}%
  \BibitemOpen
  \bibfield  {author} {\bibinfo {author} {\bibfnamefont {I.~F.}\ \bibnamefont
  {Araujo}}, \bibinfo {author} {\bibfnamefont {D.~K.}\ \bibnamefont {Park}},
  \bibinfo {author} {\bibfnamefont {F.}~\bibnamefont {Petruccione}}, \ and\
  \bibinfo {author} {\bibfnamefont {A.~J.}\ \bibnamefont {da~Silva}},\ }\href
  {\doibase 10.1038/s41598-021-85474-1} {\bibfield  {journal} {\bibinfo
  {journal} {Scientific reports}\ }\textbf {\bibinfo {volume} {11}},\ \bibinfo
  {pages} {6329} (\bibinfo {year} {2021})}\BibitemShut {NoStop}%
\bibitem [{\citenamefont {Le}\ \emph {et~al.}(2011)\citenamefont {Le},
  \citenamefont {Iliyasu}, \citenamefont {Dong},\ and\ \citenamefont
  {Hirota}}]{FRQI}%
  \BibitemOpen
  \bibfield  {author} {\bibinfo {author} {\bibfnamefont {P.}~\bibnamefont
  {Le}}, \bibinfo {author} {\bibfnamefont {A.}~\bibnamefont {Iliyasu}},
  \bibinfo {author} {\bibfnamefont {F.}~\bibnamefont {Dong}}, \ and\ \bibinfo
  {author} {\bibfnamefont {K.}~\bibnamefont {Hirota}},\ }\href {\doibase
  10.1007/s11128-010-0177-y} {\bibfield  {journal} {\bibinfo  {journal}
  {Quantum Information Processing}\ }\textbf {\bibinfo {volume} {10}},\
  \bibinfo {pages} {63} (\bibinfo {year} {2011})}\BibitemShut {NoStop}%
\bibitem [{\citenamefont {Fujii}\ and\ \citenamefont
  {Nakajima}(2021)}]{Fujii2021}%
  \BibitemOpen
  \bibfield  {author} {\bibinfo {author} {\bibfnamefont {K.}~\bibnamefont
  {Fujii}}\ and\ \bibinfo {author} {\bibfnamefont {K.}~\bibnamefont
  {Nakajima}},\ }\enquote {\bibinfo {title} {Quantum reservoir computing: A
  reservoir approach toward quantum machine learning on near-term quantum
  devices},}\ in\ \href@noop {} {\emph {\bibinfo {booktitle} {Reservoir
  Computing: Theory, Physical Implementations, and Applications}}},\ \bibinfo
  {editor} {edited by\ \bibinfo {editor} {\bibfnamefont {K.}~\bibnamefont
  {Nakajima}}\ and\ \bibinfo {editor} {\bibfnamefont {I.}~\bibnamefont
  {Fischer}}}\ (\bibinfo  {publisher} {Springer Singapore},\ \bibinfo {address}
  {Singapore},\ \bibinfo {year} {2021})\ pp.\ \bibinfo {pages}
  {423--450}\BibitemShut {NoStop}%
\bibitem [{\citenamefont {Ghosh}\ \emph {et~al.}(2019)\citenamefont {Ghosh},
  \citenamefont {Opala}, \citenamefont {Matuszewski}, \citenamefont {Paterek},\
  and\ \citenamefont {Liew}}]{QRC2}%
  \BibitemOpen
  \bibfield  {author} {\bibinfo {author} {\bibfnamefont {S.}~\bibnamefont
  {Ghosh}}, \bibinfo {author} {\bibfnamefont {A.}~\bibnamefont {Opala}},
  \bibinfo {author} {\bibfnamefont {M.}~\bibnamefont {Matuszewski}}, \bibinfo
  {author} {\bibfnamefont {T.}~\bibnamefont {Paterek}}, \ and\ \bibinfo
  {author} {\bibfnamefont {T.~C.~H.}\ \bibnamefont {Liew}},\ }\href@noop {}
  {\bibfield  {journal} {\bibinfo  {journal} {npj Quantum Information}\
  }\textbf {\bibinfo {volume} {5}},\ \bibinfo {pages} {35} (\bibinfo {year}
  {2019})}\BibitemShut {NoStop}%
\bibitem [{\citenamefont {Mujal}\ \emph {et~al.}(2021)\citenamefont {Mujal},
  \citenamefont {Martínez-Peña}, \citenamefont {Nokkala}, \citenamefont
  {García-Beni}, \citenamefont {Giorgi}, \citenamefont {Soriano},\ and\
  \citenamefont {Zambrini}}]{reviewQRC}%
  \BibitemOpen
  \bibfield  {author} {\bibinfo {author} {\bibfnamefont {P.}~\bibnamefont
  {Mujal}}, \bibinfo {author} {\bibfnamefont {R.}~\bibnamefont
  {Martínez-Peña}}, \bibinfo {author} {\bibfnamefont {J.}~\bibnamefont
  {Nokkala}}, \bibinfo {author} {\bibfnamefont {J.}~\bibnamefont
  {García-Beni}}, \bibinfo {author} {\bibfnamefont {G.~L.}\ \bibnamefont
  {Giorgi}}, \bibinfo {author} {\bibfnamefont {M.~C.}\ \bibnamefont {Soriano}},
  \ and\ \bibinfo {author} {\bibfnamefont {R.}~\bibnamefont {Zambrini}},\
  }\href@noop {} {\bibfield  {journal} {\bibinfo  {journal} {Adv. Quantum
  Technol.}\ }\textbf {\bibinfo {volume} {4}},\ \bibinfo {pages} {2100027}
  (\bibinfo {year} {2021})}\BibitemShut {NoStop}%
\bibitem [{\citenamefont {Mart\'{\i}nez-Pe\~na}\ \emph
  {et~al.}(2021)\citenamefont {Mart\'{\i}nez-Pe\~na}, \citenamefont {Giorgi},
  \citenamefont {Nokkala}, \citenamefont {Soriano},\ and\ \citenamefont
  {Zambrini}}]{DynamicalIsing}%
  \BibitemOpen
  \bibfield  {author} {\bibinfo {author} {\bibfnamefont {R.}~\bibnamefont
  {Mart\'{\i}nez-Pe\~na}}, \bibinfo {author} {\bibfnamefont {G.~L.}\
  \bibnamefont {Giorgi}}, \bibinfo {author} {\bibfnamefont {J.}~\bibnamefont
  {Nokkala}}, \bibinfo {author} {\bibfnamefont {M.~C.}\ \bibnamefont
  {Soriano}}, \ and\ \bibinfo {author} {\bibfnamefont {R.}~\bibnamefont
  {Zambrini}},\ }\href {\doibase 10.1103/PhysRevLett.127.100502} {\bibfield
  {journal} {\bibinfo  {journal} {Phys. Rev. Lett.}\ }\textbf {\bibinfo
  {volume} {127}},\ \bibinfo {pages} {100502} (\bibinfo {year}
  {2021})}\BibitemShut {NoStop}%
\bibitem [{\citenamefont {Kawai}\ and\ \citenamefont
  {Nakagawa}(2020)}]{quantumchemQRC}%
  \BibitemOpen
  \bibfield  {author} {\bibinfo {author} {\bibfnamefont {H.}~\bibnamefont
  {Kawai}}\ and\ \bibinfo {author} {\bibfnamefont {Y.}~\bibnamefont
  {Nakagawa}},\ }\href@noop {} {\bibfield  {journal} {\bibinfo  {journal}
  {Mach. Learn.: Sci. Technol.}\ }\textbf {\bibinfo {volume} {1}} (\bibinfo
  {year} {2020})}\BibitemShut {NoStop}%
\bibitem [{\citenamefont {Domingo}\ \emph {et~al.}(2022)\citenamefont
  {Domingo}, \citenamefont {Carlo},\ and\ \citenamefont {Borondo}}]{Domingo}%
  \BibitemOpen
  \bibfield  {author} {\bibinfo {author} {\bibfnamefont {L.}~\bibnamefont
  {Domingo}}, \bibinfo {author} {\bibfnamefont {G.}~\bibnamefont {Carlo}}, \
  and\ \bibinfo {author} {\bibfnamefont {F.}~\bibnamefont {Borondo}},\ }\href
  {\doibase 10.1103/PhysRevE.106.L043301} {\bibfield  {journal} {\bibinfo
  {journal} {Phys. Rev. E}\ }\textbf {\bibinfo {volume} {106}},\ \bibinfo
  {pages} {L043301} (\bibinfo {year} {2022})}\BibitemShut {NoStop}%
\bibitem [{\citenamefont {Latorre}\ and\ \citenamefont
  {Mart\'{\i}n-Delgado}(2002)}]{majorization_original}%
  \BibitemOpen
  \bibfield  {author} {\bibinfo {author} {\bibfnamefont {J.~I.}\ \bibnamefont
  {Latorre}}\ and\ \bibinfo {author} {\bibfnamefont {M.~A.}\ \bibnamefont
  {Mart\'{\i}n-Delgado}},\ }\href {\doibase 10.1103/PhysRevA.66.022305}
  {\bibfield  {journal} {\bibinfo  {journal} {Phys. Rev. A}\ }\textbf {\bibinfo
  {volume} {66}},\ \bibinfo {pages} {022305} (\bibinfo {year}
  {2002})}\BibitemShut {NoStop}%
\bibitem [{\citenamefont {Vallejos}\ \emph {et~al.}(2021)\citenamefont
  {Vallejos}, \citenamefont {de~Melo},\ and\ \citenamefont
  {Carlo}}]{majorization}%
  \BibitemOpen
  \bibfield  {author} {\bibinfo {author} {\bibfnamefont {R.}~\bibnamefont
  {Vallejos}}, \bibinfo {author} {\bibfnamefont {F.}~\bibnamefont {de~Melo}}, \
  and\ \bibinfo {author} {\bibfnamefont {G.~G.}\ \bibnamefont {Carlo}},\
  }\href@noop {} {\bibfield  {journal} {\bibinfo  {journal} {Phys. Rev. A}\
  }\textbf {\bibinfo {volume} {104}},\ \bibinfo {pages} {012602} (\bibinfo
  {year} {2021})}\BibitemShut {NoStop}%
\bibitem [{\citenamefont {Kutvonen}\ \emph {et~al.}(2020)\citenamefont
  {Kutvonen}, \citenamefont {Fujii},\ and\ \citenamefont {Sagawa}}]{OptQRC}%
  \BibitemOpen
  \bibfield  {author} {\bibinfo {author} {\bibfnamefont {A.}~\bibnamefont
  {Kutvonen}}, \bibinfo {author} {\bibfnamefont {K.}~\bibnamefont {Fujii}}, \
  and\ \bibinfo {author} {\bibfnamefont {T.}~\bibnamefont {Sagawa}},\
  }\href@noop {} {\bibfield  {journal} {\bibinfo  {journal} {Sci. Rep.}\
  }\textbf {\bibinfo {volume} {10}},\ \bibinfo {pages} {14687} (\bibinfo {year}
  {2020})}\BibitemShut {NoStop}%
\bibitem [{\citenamefont {Nielsen}\ and\ \citenamefont
  {Chuang}(2010)}]{nielsen_chuang}%
  \BibitemOpen
  \bibfield  {author} {\bibinfo {author} {\bibfnamefont {M.~A.}\ \bibnamefont
  {Nielsen}}\ and\ \bibinfo {author} {\bibfnamefont {I.~L.}\ \bibnamefont
  {Chuang}},\ }\href {\doibase 10.1017/CBO9780511976667} {\emph {\bibinfo
  {title} {Quantum Computation and Quantum Information: 10th Anniversary
  Edition}}}\ (\bibinfo  {publisher} {Cambridge University Press},\ \bibinfo
  {year} {2010})\BibitemShut {NoStop}%
\end{thebibliography}%

\end{document}